\documentclass[
superscriptaddress,
amsmath,amssymb,
aps,
prb,
twocolumn,
floatfix
]{revtex4}
\usepackage{graphicx}
\usepackage{amssymb,amsmath,bbm,braket,indentfirst,bm}
\allowdisplaybreaks
\usepackage{color}
\newcommand{\abs}[1]{\ensuremath \left|#1\right|}

\newcommand{\pri}[1]{\ensuremath #1^{\prime}}

\renewcommand\({\ensuremath \left(}
\renewcommand\){\ensuremath \right)}
\renewcommand\[{\ensuremath \left[}
\renewcommand\]{\ensuremath \right]}

\begin{document}

\title{Topological gaps without masses in driven graphene-like systems}

\author{Thomas~Iadecola}
\affiliation{Physics Department, Boston University, Boston, Massachusetts 02215, USA}

\author{Titus~Neupert}
\affiliation{Princeton Center for Theoretical Science, Princeton University, Princeton, New Jersey 08544, USA}

\author{Claudio~Chamon}
\affiliation{Physics Department, Boston University, Boston, Massachusetts 02215, USA}

\date{\today}

\begin{abstract}
We illustrate the possibility of realizing band gaps in graphene-like systems that fall outside the existing classification of gapped Dirac Hamiltonians in terms of masses.  As our primary example we consider a band gap arising due to time-dependent distortions of the honeycomb lattice.  By means of an exact, invertible, and transport-preserving mapping to a time-independent Hamiltonian, we show that the system exhibits Chern-insulating phases with quantized Hall conductivities $\pm e^2/h$.  The chirality of the corresponding gapless edge modes is controllable by both the frequency of the driving and the manner in which sublattice symmetry is broken by the dynamical lattice modulations.  Finally, we discuss a promising possible realization of this physics in photonic lattices.
\end{abstract}

\maketitle

\section{Introduction}
The critical Dirac fermions that emerge at low energies on the honeycomb lattice of graphene have inspired groundbreaking theoretical discoveries of topological phases in condensed matter physics, such as the Chern insulator \cite{haldane} and the $\mathbb{Z}_2$ topological insulator.\cite{kane1,kane2}  Various gapped phases of Dirac fermions can be realized theoretically by introducing mass terms into the Hamiltonian.\cite{masses,herbut}  For a generic Dirac Hamiltonian $\mathcal{H}_{\mathrm{D}}=p_i\alpha_i+mM$ in two spatial dimensions, the matrix $mM$ ($m\in\mathbb R$) constitutes a mass term if $M^2=\mathbbm 1$ and $\{M,\alpha_{1,2}\}=0$ for anticommuting $\alpha_i$.  This is simply because the dispersion in the presence of such a matrix takes the familiar relativistic form $E(\bm p)=\pm\sqrt{|\bm p|^2+m^2}$.  For spinless Dirac fermions in graphene, which realize a four-dimensional representation of the Dirac equation, there are four mass matrices $M$ that satisfy the necessary anticommutation relations.  More mass matrices are possible when higher-dimensional representations are considered, for example by adding further degrees of freedom, such as spin.\cite{masses}  While these masses do not occur spontaneously in graphene, there are several theoretical proposals for generating band gaps in graphene by driving it away from equilibrium.\cite{oka,kitagawa,rotatingkekule}  Although the band gaps in these proposals arise from periodic driving, they nevertheless reduce to Dirac masses in either the high-\cite{oka,kitagawa} or low-frequency\cite{rotatingkekule} limit.  

\begin{figure}[b]
\centering
\includegraphics[width=.3\textwidth]{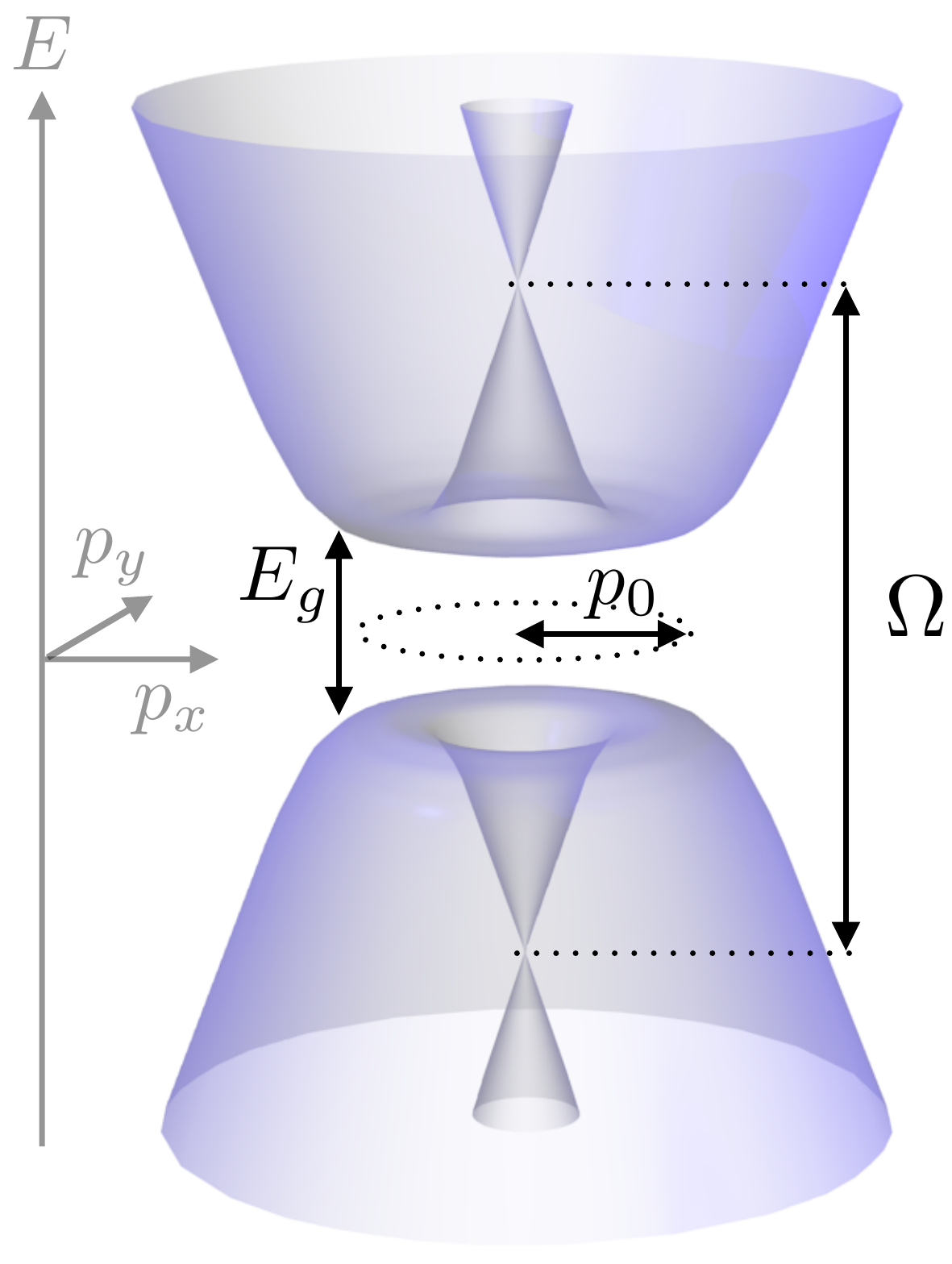}
\caption{(Color online) Schematic of the low-energy band structure of the Hamiltonian \eqref{tindep} with $M(\bm p)$ given by Eq.~\eqref{pdm}.  When only one sublattice is excited  (i.e.\ $\Delta_B=0$), the gap opens along a circle of radius $p_0$ in momentum space. }\label{bands}
\end{figure}

In this paper, we provide an example of a band gap in graphene that does not reduce to a Dirac mass in any limit.  To do this, we consider a dynamical coupling of the two Dirac points, which can be generated by driving a superposition of phonon modes in the honeycomb lattice.  By virtue of a time-dependent gauge transformation, the driven Hamiltonian maps bijectively to a time-independent Bloch Hamiltonian. The latter features two bands with an avoided crossing along a circle (or, more generally, an ellipse) of radius (major axis) $\sim\Omega/2$, where $\Omega$ is the frequency of the driven modes (see Fig.~\ref{bands}).  (We employ units where the Fermi velocity $v_F=\hbar=1$ unless otherwise noted.)  Depending on which phonon mode is exited, this gap can have either $s$- or (chiral) $d$-wave character.  This goes markedly beyond the classification of Dirac masses, which are always $s$-wave by definition.  While gaps with nontrivial angular momentum occur generically in the particle-hole symmetric Bogoliubov bands of mean-field superconductors, where the superconducting order parameter can carry arbitrary angular momentum in $\mathbb Z$, this scenario is less common in noninteracting electronic bands.

%The mass gaps lift an accidental (or symmetry-protected) touching of two bands at a single \emph{point} in momentum space, where the lifting is associated with a $\mathbb{Z}_2$ degree of freedom at each such point, namely the sign of the mass.
%One would anticipate a much richer structure if instead the degeneracy-lifting occurred along a closed one-dimensional \emph{curve} in momentum space. The gaps can then carry a nontrivial angular momentum structure. While such one-dimensional degeneracies occur generically in the particle-hole symmetric Bogoliubov bands of mean-field superconductors, where the order parameter of arbitrary angular momentum in $\mathbb{Z}$ lifts the degeneracy, this scenario is less common to noninteracting electronic bands. [References?]

We exemplify this mechanism by considering a particular superposition of phonon modes that modifies the single-particle spectrum by modulating both the onsite potential and the bond lengths in the lattice. The time-independent Hamiltonian obtained after the gauge transformation is fully gapped and belongs to either symmetry class C or A,\cite{az,schnyder} depending on whether the onsite potential or bond length modulation dominates.  In both cases, the phonon modes break time-reversal symmetry and the system supports a Chern number $C=\pm1$.  It turns out that the time-dependent gauge transformation leaves the electromagnetic current invariant, and we conclude that the steady state of the driven system features a quantized Hall conductivity $\sigma_{xy}=C\,e^2/h$.  Due to the exact nature of the mapping, these results do not rely on any of the approximations that are employed in the Floquet treatment of driven systems.\cite{shirley,sambe,oka,kitagawa}  We further show that while this model is topologically equivalent to the Chern insulator constructed by Haldane,\cite{haldane} it is nevertheless separated from the latter by a gap-closing transition in a certain parameter regime.  We close by proposing a realization of this physics in photonic lattices.

\section{Archetypal Model}
We consider spinless, noninteracting fermions hopping on a hexagonal lattice $\Lambda=\Lambda_A\cup\Lambda_B$, where $\Lambda_{A,B}$ are interpenetrating triangular sublattices.  The lattice is subjected to time-dependent in-plane deformations, so that the tight-binding Hamiltonian is of the form $H(t)=H_{\rm NN}(t)+H_{\rm NNN}(t)+H_{\rm OS}(t)$, where $t$ is time.  The contribution from nearest-neighbor (NN) hopping is
\begin{align}\label{hnn}
H_{\rm NN}(t)=-\sum_{\bm r\in\Lambda_A}\sum_{i=1}^3\[t_0+\delta t_{\bm r,i}(t)\] a_{\bm r}^\dagger b_{\bm r+\bm  s_i}+ \text{H.c.}
\end{align}
Here, $a_{\bm r}^\dagger$ and $b^\dagger_{\bm r+\bm s_i}$ create fermions on sublattices $A$ and $B$, respectively, and $\bm s_i$ are vectors connecting nearest-neighbor sites.  In the absence of the hopping modulation $\delta t_{\bm r,i}(t)$, $H_{\rm NN}$ has two inequivalent Dirac points located at opposite corners of the Brillouin zone, $\bm K_{\pm}=\frac{4\pi}{3\sqrt 3 d}(\pm 1,0)$, with $d\equiv |\bm s_i|$ the NN distance.  The contribution from next-nearest-neighbor (NNN) hopping is given by
\begin{align}\label{hnnn}
\hspace{-.25cm}H_{\rm NNN}(t)&=-\sum_{\bm r\in\Lambda_A}\sum_{j=1}^6\Big[ \delta t^A_{\bm r,j}(t)\; a_{\bm r}^\dagger a_{\bm r+\bm a_{j}}\nonumber\\
&\hspace{1cm}  +\; \delta t^B_{\bm r,j}(t)\; b_{\bm r+\bm s_1}^\dagger b_{\bm r+\bm s_1+\bm a_{j}}+\text{H.c.} \Big] ,
\end{align}
where for simplicity we have assumed that NNN hoppings are absent without the time-dependent lattice distortions.\footnote{Ignoring the bare NNN hopping integral $t_1$ is justified as it does not open a gap in the electronic spectrum.\cite{graphenermp}}  Here, $\bm a_j$ are vectors connecting next-nearest-neighbors, and $|\bm a_j|=\sqrt 3\; d$.  Finally, a modulation of the onsite potential is also possible:
\begin{align}\label{hos}
H_{\rm OS}(t)&=\sum_{\bm r\in\Lambda_A}\Big[\varepsilon_{AB}(\bm r,t)\; a^\dagger_{\bm r}a_{\bm r}\nonumber\\
&\hspace{1cm}+\; \varepsilon_{BA}(\bm r+\bm s_1,t)\; b^\dagger_{\bm r+\bm s_1}b_{\bm r+\bm s_1}\Big].
\end{align}
Here, $\varepsilon_{AB}$ is the potential modulation on sublattice $A$ due to the motion of sublattice $B$, and likewise for $\varepsilon_{BA}$.  Perturbations of this type are especially relevant in graphene, where the couplings entering the onsite potential modulations can be larger than those governing changes in NNN and possibly even NN hopping.\cite{basko,dft,fernando}

In this work, we consider lattice distortions with a spatial periodicity defined by the wavevectors $\bm K_{\pm}$.  At low energies, such distortions scatter fermions between the two Dirac points, which are separated by a vector $\bm G = \bm K_+-\bm K_-$.  As such, it is natural to study the band structure of $H(t)$ by transforming the creation and annihilation operators to momentum space and restrict it to the modes that are a small momentum $\bm{p}$ away from $\bm K_{\pm}$.  Such distortions lead to a Hamiltonian of the form $H(t)=\sum_{\bm p}\Psi_{\bm p}^\dagger\mathcal H_{\bm p}(t)\Psi_{\bm p}$, where $\Psi_{\bm p}^\dagger=(a_{+,\bm p}^\dagger\;b_{+,\bm p}^\dagger\;b_{-,\bm p}^\dagger\;a_{-,\bm p}^\dagger)$ combines the creation operators for fermions of momentum $\bm K_{\pm}+\bm p$.  The single-particle Hamiltonian $\mathcal H_{\bm p}(t)$ can then be written in block form, to leading order in $\bm p$, as 
\begin{align}\label{generic}
\mathcal H_{\bm p}(t)=\begin{pmatrix}
\bm\sigma\cdot\bm p & M(\bm p,t)\\
M^\dagger(\bm p,t)&-\bm\sigma\cdot\bm p
\end{pmatrix}.
\end{align}
Here, $\bm \sigma = (\sigma_1,\sigma_2)$ is the vector of Pauli matrices that act on sublattice indices, and $M(\bm p,t)$ is a $2\times 2$ matrix whose form depends on the details of the distortions.

%Eq. \eqref{generic} defines a Dirac Hamiltonian with arbitrary off-diagonal time-dependence.  In general, the analysis of such time-dependent Hamiltonians is nontrivial, as the lack of time-translation invariance makes the notion of energy levels ill-defined. 

If the time-dependent lattice distortion of frequency $\Omega$ scatters fermions between $\bm K_{+}$ and $\bm K_{-}$, $M(\bm p,t)$ takes the form $M(\bm p,t)=M(\bm p)\; e^{-i\Omega t}$. 
In this case, the Hamiltonian \eqref{generic} can be brought to a time-independent form by making an appropriate unitary transformation to a ``rotating frame," as we now show.  The eigenstates $\ket{\Psi_{\bm p}}$ of $\mathcal H_{\bm p}(t)$ satisfy the time-dependent Schr\"odinger equation $\mathcal{H}_{\bm p}(t)\ket{\Psi_{\bm p}}=i\partial_t\ket{\Psi_{\bm p}}$. Let us now define $\ket{\Psi_{\bm p}}=U(t)\ket{\tilde\Psi_{\bm p}}$, with $U(t)=\exp(i\gamma_5\Omega t/2)$, where $\gamma_5\equiv\tau_3\otimes\sigma_0$, $\sigma_0$ is the identity matrix in sublattice space, and $\tau_3$ is a Pauli matrix acting in valley space.  One can then rewrite the original Schr\"odinger equation as $\tilde{\mathcal H}_{\bm p}\ket{\tilde\Psi_{\bm p}}=i\partial_t\ket{\tilde\Psi_{\bm p}}$, with
\begin{align}
\tilde{\mathcal H}_{\bm p} &= U(t)\mathcal H_{\bm p}U^\dagger(t)-iU(t)\partial_tU^\dagger(t)\nonumber\\
&=\begin{pmatrix}
\bm\sigma\cdot\bm p-\frac{\Omega}{2}\; \sigma_0 & M(\bm p)\\
M^\dagger(\bm p)&-\bm\sigma\cdot\bm p+\frac{\Omega}{2}\; \sigma_0
\end{pmatrix}.\label{tindep}
\end{align}

The band structure of the time-independent Hamiltonian \eqref{tindep} characterizes the physical properties of the driven system in a sense that we will now make precise.  First, we note that the transformation $U(t)$ preserves the $U(1)$ current density $j^l_{\bm p}=\bra{\Psi_{\bm p}}\gamma_0\gamma^l\ket{\Psi_{\bm p}}$, where $l\in\{1,2\}$, $\gamma_0=\tau_1\otimes\sigma_0$, and $\gamma^l=-i\; \tau_2\otimes\sigma_l$.  Consequently, all electronic transport properties (e.g.\ conductivities) of the driven system are captured by the time-independent Hamiltonian $\tilde{\mathcal H}_{\bm p}$.\cite{rotatingkekule}   Furthermore, one can couple the system to a heat bath of acoustic phonons and carry out the procedure of Ref.\ \onlinecite{floquetkekule}, which demonstrates a bijection between the steady-state occupation numbers of the driven system and the equilibrium occupation numbers of the corresponding time-independent Hamiltonian.  As a result, many relevant questions regarding the nature of the electronic steady state of the driven system can be answered by studying the time-independent Hamiltonian \eqref{tindep}.  For example, one can classify these steady states according to the discrete symmetries of $\tilde{\mathcal H}_{\bm p}$.  Furthermore, the invariance of the current density allows one to use the bulk-boundary correspondence for $\tilde{\mathcal H}_{\bm p}$ to diagnose the topological sector within each symmetry class, exactly as in the time-independent scenario.\cite{az,schnyder}

In Ref.\ \onlinecite{rotatingkekule} it is shown that exciting the TO (transverse optical) phonon mode of graphene at momentum $\bm k=\bm K_{\pm}$ leads to a Hamiltonian of the form \eqref{tindep} in the rotating frame, with $M(\bm p)=\Delta\; \sigma_0$ and $\Delta\in\mathbb C$.  The spectrum of $\tilde{\mathcal H}_{\bm p}$ is given by $E_{\pm,\mp}(\bm p)=\pm\sqrt{(|\bm p|\mp\Omega/2)^2+|\Delta|^2}$, indicating the presence of a gap of size $2|\Delta|$ in the driven system.  The degeneracy of the valence and conduction bands is lifted along a circle of radius $\Omega/2$ in momentum space, rather than at a single Dirac point, but in the adiabatic limit $\Omega\to 0$, $E_{\pm,\mp}\to\pm\sqrt{|\bm p|^2+|\Delta|^2}$, which is of the usual massive Dirac form.  The TO phonon therefore realizes a dynamical version of the so-called Kekul\'e mass. \cite{masses}  We will now study two examples where the resulting gap \emph{does not} reduce to a mass gap in the adiabatic limit.

\section{Case Study: LO/LA Phonon Modes in Graphene}
We consider a superposition of two graphene phonon modes, known as the LA (longitudinal acoustic) and LO (longitudinal optical) modes.  We again excite the modes with momentum $\bm k=\bm K_+$,\footnote{The analysis of this paper could also be carried out for $\bm k=\bm K_-$, in which case $\Omega\to -\Omega$.} for which they are degenerate with energy $\Omega\sim 150$ meV.\cite{svkphonons}  Together the LA and LO modes modulate sublattices $A$ and $B$ independently,\cite{ando} and the displacements of the atoms from their equilibrium positions can be written as
\begin{subequations}\label{modes}
\begin{align}
u^A_+(\bm r_A,t)&= \sqrt 2\;  c_+^{A*}\; e^{-i\bm r_A\cdot\bm K_+}e^{i\Omega t}\\
u^B_+(\bm r_B,t)&=- \sqrt 2\;  c_+^{B}\; e^{i\bm r_B\cdot\bm K_+}e^{-i\Omega t},
\end{align}
\end{subequations}
where $c_+^{A,B}=c_{+,x}^{A,B}+i\,c_{+,y}^{A,B}$ are complex amplitudes for the mode in either sublattice.  Below we will focus separately on two effects arising due to the lattice distortions of Eqs.~\eqref{modes}, namely the onsite potential modulation and the hopping modulations mentioned above.

\subsection{Onsite potential}
We turn first to the change in onsite potential, which is the dominant effect in graphene when the LO/LA modes are excited at $\bm K_\pm$.\cite{dft,fernando}  To leading order in displacements, the potential variation in sublattice $A$ [c.f.\ Eq.\ \eqref{hos}] is given by
\begin{align}
\varepsilon_{AB}(\bm r,t)&\approx\frac{\varepsilon}{3}\sum_{j=1}^3iz_j^*\[u^A_+(\bm r)-u^B_+(\bm r+\bm s_j)\]+\text{c.c.},
\end{align}
where $z_j=e^{i\; 2\pi(j-1)/3}$ are cubic roots of unity.  DFT estimates\cite{dft} suggest that the coupling $\varepsilon\approx -6\ \text{eV/\AA}$ (relative to an appropriately chosen vacuum energy).  Evaluating this expression using the sum rule $\sum_{j=1}^3z_j=0$, we find that
\begin{align}\label{epsab}
\varepsilon_{AB}(\bm r,t)=i\; \alpha_B\; e^{-i\bm G\cdot\bm r}\; e^{-i\Omega t}+\text{c.c.},
\end{align}
where $\alpha_B=\sqrt 2\; \varepsilon\; c_+^B$.  A similar calculation for sublattice $B$ yields
\begin{align}\label{epsba}
\varepsilon_{BA}(\bm r,t)=-i\; \alpha_A\; e^{-i\bm G\cdot\bm r}\; e^{-i\Omega t}+\text{c.c.},
\end{align}
where $\alpha_A=\sqrt 2\; \varepsilon\; c_+^A$.  The single-particle Hamiltonian is obtained by substituting the potential modulations~\eqref{epsab} and~\eqref{epsba} into Eq.~\eqref{hos} and working to leading order in momenta near the Dirac points.  Performing the gauge transformation $U(t)$ to remove the time-dependence, we obtain a Hamiltonian $\tilde{\mathcal H}_{\bm p}$ of the form \eqref{tindep}, with
\begin{align}\label{onsitem}
M(\bm p)=M=\begin{pmatrix}
0&i\; \alpha_B\\
-i\; \alpha_A&0
\end{pmatrix}.
\end{align}

Although the matrix structure of Eq.~\eqref{onsitem} precludes its interpretation as a mass term in the Hamiltonian \eqref{tindep}, it nevertheless opens a gap at finite momentum, as we now show.  The spectrum of the Hamiltonian \eqref{tindep} with $M(\bm p)=M$ as in Eq.~\eqref{onsitem} can be found exactly for arbitrary $\alpha_{A,B}$.  For the simple case where only one sublattice is excited ($\alpha_B=0$, say), it takes the form
\begin{align}\label{onsitespec}
E_{\pm,\mp}&=\pm \sqrt{ p^2+\frac{\Omega^2}{4}+\frac{\alpha_A^2}{2}\mp\frac{1}{2} \sqrt{\alpha_A^4+4 p^2 \left(\alpha_A^2+\Omega ^2\right)}},
\end{align}
where $p=|\bm p|$.  The size of the gap in this case is
\begin{align}\label{onsitegap}
\min_{p}\(E_{+,-}-E_{-,-}\)\equiv E_{\rm g}=\frac{\alpha_A\; \Omega}{\sqrt{\alpha_A^2+\Omega^2}},
\end{align}
with the minimum occurring along a circle in momentum space of radius
\begin{align}
p_{\rm min}=\frac{\Omega}{2}\sqrt{1+\frac{\alpha_A^2}{\alpha_A^2+\Omega^2}}.
\end{align}
It is crucial to note that the gap \eqref{onsitegap} scales with $\Omega$ and therefore vanishes in the absence of driving.  The situation is identical if we instead choose $\alpha_A=0$ and $\alpha_B\neq 0$.  If both $\alpha_A$ and $\alpha_B$ are nonzero, corresponding to the case where both degenerate phonon modes are excited, then the system remains gapped unless $|\alpha_A|=|\alpha_B|$, in which case the system recovers sublattice symmetry because both of the LO/LA modes are excited with equal amplitude.

\subsection{Hopping modulation}
While the onsite potential is the dominant effect arising due to the LO/LA modes in graphene, there are also graphene-like systems where the lattice sites do not carry charge and therefore do not modulate the onsite potential when in motion.  (This is the case, for example, in the photonic lattices discussed in Section IV.)  In this case, the hopping modulations arising from the time-periodic distortions of the bond lengths are the dominant effect.  The change in NN bond lengths due to the phonon modes is, to leading order in the displacements, given by \cite{solitons}
\begin{align}
\frac{\delta d_{\bm r,j}(t)}{d}\approx-\frac{i}{2}\frac{z_j^*}{d}\[u_{+}^A(\bm r)-u_{+}^B(\bm r+\bm s_j)\]+\text{c.c.}\label{nnbond}.
\end{align}
These changes in bond lengths induce a NN hopping distortion
\begin{align}
\delta t_{\bm r,j}(t)\approx-\frac{ig}{\sqrt 2}\(c_+^{A}z_j-c_+^{B}\)\; e^{-i\Omega t} e^{-i\bm G\cdot\bm r}+{\text{c.c.}}.\label{nn}
\end{align}
The electron-phonon coupling $g$ has been estimated\cite{dft} to be between $4.5$ and $7.8\ \text{eV/\AA}$.    Similarly, the change in NNN bond lengths for sublattice $A$ is
\begin{align}
\frac{\delta d^{A}_{\bm r,j}(t)}{d\sqrt 3}&\approx -\frac{1}{2}\frac{w_j^*}{d\sqrt 3}\[u_{+}^{A}(\bm r,t)- u_{+}^{A}(\bm r+\bm a_k,t)\]+\text{c.c.},
\end{align}
where $w_j=e^{i\; 2\pi(j-1)/6}$ are sixth roots of unity.  The resulting NNN hopping distortion is
\begin{align}
\delta t^A_{\bm r,j}(t)\approx\tau\frac{g}{\sqrt 6}\; c_+^A\;  w_j(1-e^{i\bm K_+\cdot \bm a_j})\; e^{-i\Omega t}e^{-i\bm G\cdot\bm r}+\text{c.c.},\label{annn}
\end{align}
where $\tau\equiv t_1/t_0$, with $t_1$ the bare NNN hopping integral.  A recent measurement indicates that \nolinebreak{${\tau\sim 1/10}$}.~\cite{nnnmeasurement}  A similar construction determines the NNN hopping on sublattice B.  Substituting the hopping modulations~\eqref{nn} and~\eqref{annn} into Eqs.~\eqref{hnn} and~\eqref{hnnn}, we obtain a Hamiltonian $\tilde{\mathcal H}_{\bm p}$ in the rotating frame of the form \eqref{tindep}, with
\begin{align}\label{pdm}
M(\bm p)=\begin{pmatrix}
\Delta_A\; p^*+\Delta_B^*\;  p&-6\tau d\Delta_A\; (p^*)^2\\
6\tau d\Delta_B^*\; p^2&-(\Delta_A\; p^*+\Delta_B^*\;  p)
\end{pmatrix},
\end{align}
where $p\equiv p_x+i p_y$.  Note that $\Delta_A\propto t_0 \; c_+^A$ and $\Delta_B^*\propto t_0\; c_+^B$ up to constants of order unity.

%\begin{figure}
%\centering
%\includegraphics[width=.3\textwidth]{bands}
%\caption{Schematic of the low-energy band structure of the Hamiltonian \eqref{tindep} with $M(\bm p)$ given by \eqref{pdm}.  When only one sublattice is excited  (i.e.\ $\Delta_B=0$), the gap opens along a circle of radius $p_0$ in momentum space. }
%\end{figure}

%\input{gap.tex}

%\begin{figure}[t]
%\centering
%(a)\includegraphics[width=.23\textwidth]{d-phase_cropped}
%\hspace{-4mm}(b)\includegraphics[width=.22\textwidth]{haldane-d_cropped}
%\caption{
%(a) Phase diagram of the $d$-wave model.  Chern-insulating regions with $C=\pm 1$ are separated by time-reversal-symmetric gap-closing transitions along the lines $|\Delta_B|=\pm|\Delta_A|$, i.e.\ when the phonon modes of Eqs.\ \eqref{modes} are excited with equal amplitudes.
%(b) Phase diagram for the Hamiltonian $\mathcal H_{\eta\Omega}$.  When $\alpha=0$, the $d$-wave and Haldane phases are separated by a gap-closing transition.  However, when $\alpha\neq0$, this critical point can be avoided and the two phases can be connected without closing the gap.
%}
%\end{figure}

We now investigate the effect of these hopping modulations on the single-particle spectrum.  Let us again first study the case where only one of the degenerate phonon modes is excited at a time.  Adopting polar coordinates $p=|p|e^{i\theta}$, we consider the case where only $\Lambda_A$ is excited (i.e.\ $\Delta_B=0$), although the following  holds equally well if $\Lambda_B$ is excited.  Performing successive $SU(2)\times SU(2)$ rotations to remove the phases of $\Delta_A$ and $p$, we find the following simplified Hamiltonian:
\begin{align}
\tilde{\mathcal H}_{\bm p}=\begin{pmatrix}
-\Omega/2&|p|&\Delta |p|&-6d\tau\Delta |p|^2\\
|p|&-\Omega/2&0&-\Delta |p|\\ \Delta |p|&0&\Omega/2&-|p|\\ -6d\tau\Delta |p|^2&-\Delta |p|&-|p|&\Omega/2
\end{pmatrix}\label{simplea},
\end{align}
where $\Delta\equiv |\Delta_A|$.  The spectrum of Hamiltonian~\eqref{simplea} can be obtained within degenerate perturbation theory by noting that the matrix elements of order $|p|^2$ are suppressed by a factor $d\tau$ in addition to the extra power of momentum.  In the absence of this perturbation, the Hamiltonian is gapless along the circle in momentum space of radius
\begin{align}
|p|=p_0\equiv\frac{\Omega}{2}\frac{1}{\sqrt{1-\Delta^2}}\label{p0}.
\end{align}
Introducing the perturbation of order $|p|^2$ lifts the degeneracy along this circle. Taking matrix elements of the perturbation with the eigenstates of the unperturbed Hamiltonian and projecting onto the low-energy sector, we obtain the following estimate (which is essentially exact for $\Delta^2\ll1$) of the size of the gap:
\begin{align}\label{gap}
E_{\rm g}\approx \frac{3\tau}{2}\; \frac{d\Delta}{1-\Delta^2}\; \Omega^2.
\end{align}
We note that the gap described above is of dominantly $d$-wave character in the following sense.  Consider deforming $\tilde{\mathcal H}_{\bm p}$ by sending $\Delta\to 0$ and $d\to \infty$ in such a way that $d\Delta\equiv\text{const}$.  Then $E_{\rm g}\propto d\Delta\Omega^2$, which vanishes only in the absence of driving.  In other words, turning off the $p$-wave portion of $M(\bm p)$ [i.e.\ the diagonal entries in Eq.~\eqref{pdm}], does not close the gap, and so the $d$-wave portion [i.e.\ the off-diagonal entries in Eq.~\eqref{pdm}] instead controls the transport properties of the steady state.  In this sense, it is appropriate to refer to the Hamiltonian $\tilde{\mathcal H}_{\bm p}$ with $M(\bm p)$ given by Eq.\ \eqref{pdm} as the $d$-wave model.  This observation also serves as a justification \emph{ex post facto} of our neglect of the NN hopping when we studied the onsite potential in the previous subsection; it is the $d$-wave part of the Hamiltonian, which is suppressed relative to the onsite potential by a factor of $d\tau$, that opens the gap in this case, so the two effects do not compete significantly in graphene.

\subsection{Topological properties and connection to the Haldane model}
We now turn to the topological characteristics of the low-energy spectrum of graphene in the presence of the LO/LA modes.  For simplicity we again consider the case where only sublattice $A$ is excited (i.e.~$c_+^B=0$), and we further examine separately the effects of the onsite potential (Sec.~III.A) and NNN hopping (Sec.~III.B) modulations.    To determine what topological classification is possible, we first consider whether the Hamiltonian possesses the discrete symmetries $\mathcal T$ (time reversal), $\mathcal P$ (particle-hole) and $\mathcal S$ (sublattice).  The anti-unitary symmetries are defined in terms of $\mathcal T=\mathcal KU_{\mathcal T}$ and $\mathcal P=\mathcal KU_{\mathcal P}$, where $U_{\mathcal T,\mathcal P}$ are unitary and $\mathcal K$ is complex conjugation.  The Hamiltonian $\tilde{\mathcal H}_{\bm p}$ possesses one of these symmetries if any of the following relations is satisfied:
\begin{subequations}\label{symms}
\begin{align}
U_{\mathcal T}^\dagger\tilde{\mathcal H}^*_{-\bm p}U_{\mathcal T}&=+\tilde{\mathcal H}_{\bm p}\label{trs}\\
U_{\mathcal P}^\dagger\tilde{\mathcal H}^*_{-\bm p}U_{\mathcal P}&=-\tilde{\mathcal H}_{\bm p}\label{phs}\\
\mathcal S^\dagger\tilde{\mathcal H}_{\bm p}\mathcal S&=-\tilde{\mathcal H}_{\bm p}\label{sls}.
\end{align}
\end{subequations}
The onsite potential and NNN hopping modulations fall into different symmetry classes\cite{az} with compatible topological classifications.  For the onsite potential modulation, the spectrum \eqref{onsitespec} is manifestly particle-hole-symmetric, and we consequently find that PHS is implemented by $U_{\mathcal P}=\tau_1\otimes\sigma_2$, so that $\mathcal P^2=-1$.  We find no $U_{\mathcal T}$ or $\mathcal S$ satisfying Eqs.~\eqref{trs} and \eqref{sls}, so the Hamiltonian belongs to symmetry class C.  For the hopping modulations, we find no suitable choices of $U_{\mathcal T,\mathcal P}$ or $\mathcal S$, so none of Eqs.~\eqref{symms} holds and the Hamiltonian is in class A.  In both cases, TRS is broken and the system supports a nonzero Chern number.\cite{schnyder}  If we place the Fermi energy in the gap near $E=0$, then the Chern number is given by \cite{tknn}
\begin{align}\label{chern}
C=\frac{1}{2\pi i}\sum_{\gamma=1,2}\int d^2k\; \epsilon^{ij}\; \partial_{\bm k_i}\bra{\Psi_\gamma}\partial_{\bm k_j}\ket{\Psi_\gamma},
\end{align}
where $\gamma$ labels the bands in order of increasing energy and the integral is taken over the whole plane.  It is important to note that if the two lower bands of $\tilde{\mathcal H}_{\bm p}$ are degenerate (as occurs at $|\bm p|=0$ for the NNN hopping modulation), the Chern number of an individual band is not well defined.\cite{bernevig}  Nevertheless, the Chern number for the whole system at half-filling, which is given by the sum of the Chern numbers for the two occupied bands, must be quantized.  

\begin{figure}[t]
\centering
(a)\includegraphics[width=.2\textwidth]{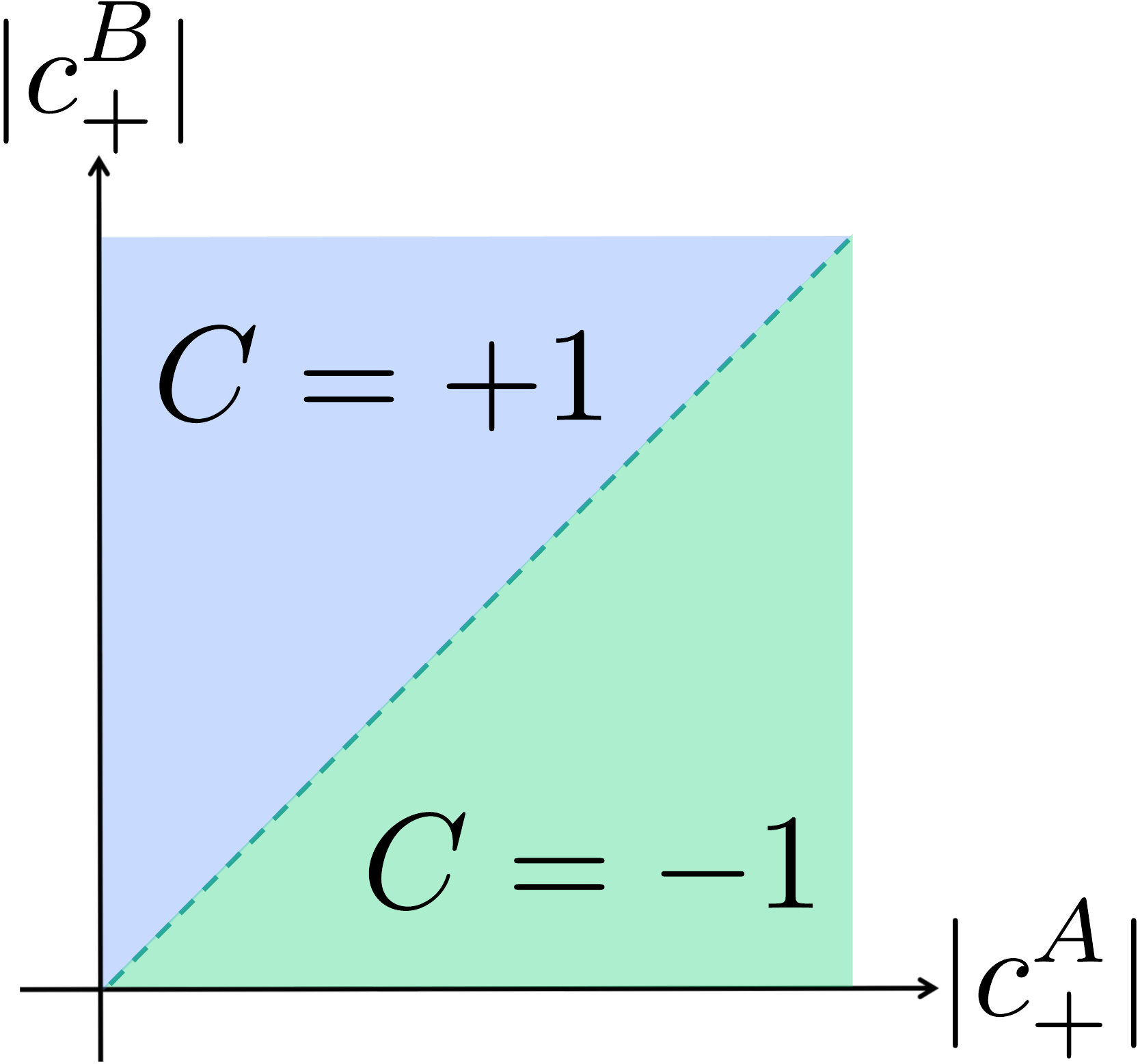}
\hspace{-.25mm}(b)\includegraphics[width=.22\textwidth]{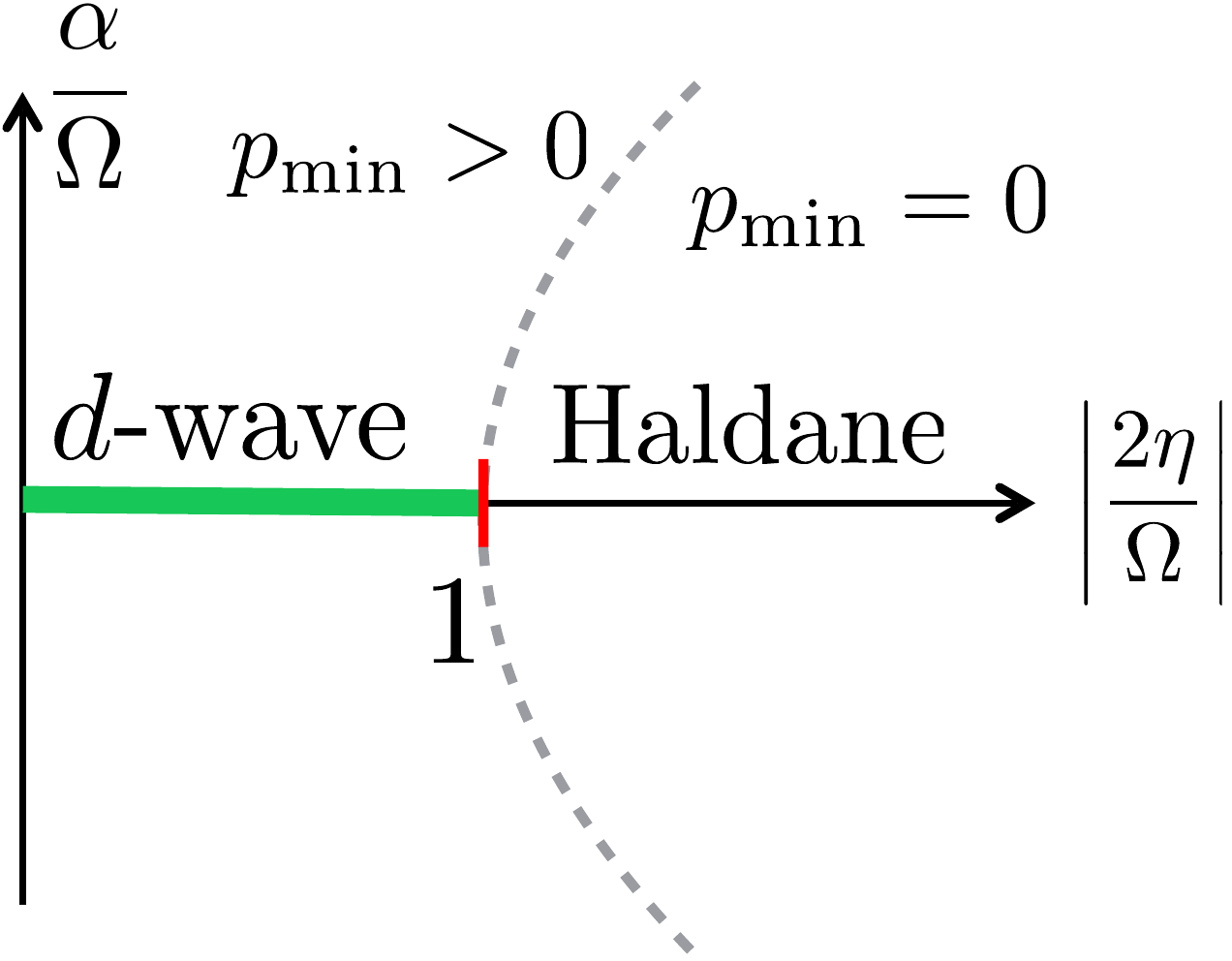}
\caption{
(Color online) 
(a) Phase diagram of graphene in the presence of the LO/LA phonon modes.  Chern-insulating regions with $C=\pm 1$ are separated by a gap-closing transition along the line $|c^B_+|=|c^A_+|$, i.e., when the phonon modes of Eqs.\ \eqref{modes} are excited with equal amplitudes.
(b) Schematic phase diagram for the Hamiltonian $\mathcal H_{\eta\Omega,\bm p}$ of Eq.~\eqref{heta}.  When $\alpha=0$, the $d$-wave and Haldane phases are separated by a gap-closing transition at $2\eta/\Omega=\pm 1$.  However, when $\alpha\neq0$, this critical point can be avoided and the two phases can be connected without closing the gap.  To the right of the dashed line, the location $p_{\rm min}$ of the minimum gap size is identically zero, while to the left $p_{\rm min}>0$.
}
\label{fig: phase diagrams}
\end{figure}

From Eq.~\eqref{chern} we calculate that, for both cases, \nolinebreak{$C=-1$} when only sublattice $A$ is excited.  In other words, the onsite potential and NNN hopping modulations lead to the same topological characteristics at half-filling.  These characteristics themselves are determined solely by parameters that are common to the two mechanisms, namely the relative amplitude of the phonon modes in either sublattice.  For example, if instead only sublattice $B$ is excited (i.e.~$c_+^A=0$), a similar calculation yields $C=+1$ in both cases.  The change in $\text{sgn } C$ must be accompanied by a gap-closing transition as one varies $|c_+^B|/|c_+^A|$ from $0$ to $\infty$.  Indeed, one verifies from Eqs.~\eqref{onsitem} and \eqref{pdm} that the gap closes when $|c_+^B|/|c_+^A|= 1$.  This gap-closing can be understood as follows.  When $|c_+^A|=|c_+^B|$, the two sublattices are excited with equal amplitudes, and this sublattice symmetry manifests itself as an effective time-reversal symmetry satisfying $\mathcal T^2=-1$ in the time-independent Hamiltonian $\tilde{\mathcal H}_{\bm p}$.  This odd TRS forces the Hamiltonian into class CII for the onsite potential modulation, or class AII for the hopping modulations.  The existence of TRS in this special case forces the Chern number to vanish.  The fact that the Chern number is constant away from this critical line allows one to construct a topological phase diagram for graphene in the presence of the LO/LA modes [see Fig.~\ref{fig: phase diagrams} (a)].  

Two comments are in order.  First, it is interesting to note that $\text{sgn }C$ is also linked to $\text{sgn }\Omega$, which is itself determined by whether the phonon modes are excited at $\bm K_+$ or $\bm K_-$.  In other words, if one had instead chosen to consider the phonon modes at $\bm K_-$, one would find $C=+1$ for the $\Lambda_A$ case and $C=-1$ for the $\Lambda_B$ case.  Second, it is worth reiterating that the Chern number calculated from $\tilde{\mathcal H}_{\bm p}$ {\it must} characterize the Hall conductivity of the driven system with its full time-dependent Hamiltonian.  This is a necessary consequence of the fact that the electromagnetic current density is invariant under the gauge transformation $U(t)$ that removes the time-dependence.  In other words, the case study considered here constitutes a dynamical realization of a Chern insulating phase with $\sigma_{xy}=\pm e^2/h$.

\begin{figure*}
\centering
\includegraphics[width=.67\textwidth]{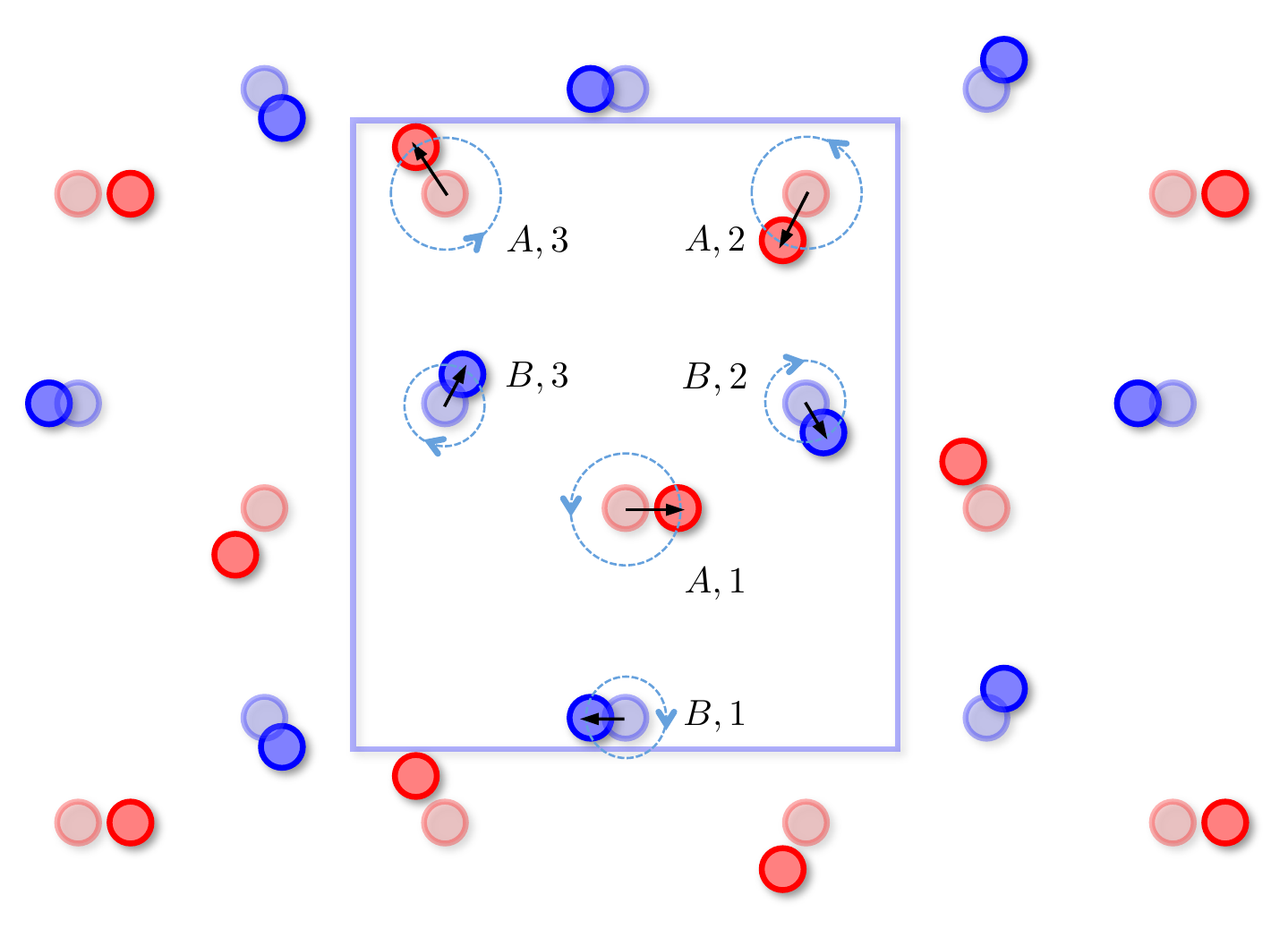}
\caption{(Color online) Motion of the honeycomb lattice in the asymmetric case where the $A$ and $B$ sublattices are excited with different amplitudes.  The faint circles indicate the equilibrium positions of the ions, around which the driven ions rotate along the dashed circles with chirality indicated by the blue arrows.  The boxed region indicates the tripled unit cell of the driven system, and the black arrows indicate the initial phases $-\phi_i$ of the inequivalent lattice sites labeled by $i=1,2,3$.}\label{dfiber}
\end{figure*}

%\begin{figure}[t]
%\centering
%\includegraphics[width=.4\textwidth]{haldane-d_cropped}
%\caption{Phase diagram for the Hamiltonian $\mathcal H_{\eta\Omega}$.  When $\alpha=0$, the $d$-wave and Haldane phases are separated by a gap-closing transition.  However, when $\alpha\neq0$, this critical point can be avoided and the two phases can be connected without closing the gap.}
%\end{figure}

The topological picture outlined above is reminiscent of the phase diagram of the Haldane model.\cite{haldane}  In the low-energy limit, the Haldane model is described by the massive Dirac Hamiltonian $\mathcal{H}_{\mathrm{H},{\bm p}}=p_i\; \tau_3\otimes\sigma_i+ \eta\; \tau_0\otimes\sigma_3$.  At half-filling, the system exhibits Chern insulating phases with $C=\text{sgn }\eta$ separated by a time-reversal symmetric gap-closing point at $\eta=0$, where we recover a massless Dirac Hamiltonian.  This similarity is not accidental---the model studied in this paper is topologically equivalent to the Haldane model, as there exists a series of continuous deformations taking one into the other without closing the gap.  To see this, consider the Hamiltonian
\begin{align}\label{heta}
\mathcal H_{\eta\Omega,{\bm p}}=\tilde{\mathcal H}_{\bm p}+\frac{\alpha}{2}\; (\tau_2\otimes\sigma_1-\tau_1\otimes\sigma_2)-\eta\; \tau_0\otimes\sigma_3,
\end{align} with $\tilde{\mathcal H}_{\bm p}$ as given in Eq.\ \eqref{simplea} and $\alpha,\ \eta>0\in\mathbb R$.  The $\alpha$ term above corresponds to an onsite potential modulation with $|\alpha_A|\equiv \alpha$.  The Hamiltonian $\mathcal H_{\eta\Omega,{\bm p}}$ exhibits two gapped phases, each with $C=-1$ at half-filling.  In one phase, the gap opens at $p=0$ (as in the Haldane model), while in the other phase the gap opens at finite momentum $|p|\sim p_{\rm min}$, as in Secs.~III.A and III.B.  If $\alpha\neq 0$, as is the case in graphene, then one can simply tune $\eta$ to move from one phase to the other without closing the gap.  If instead $\alpha=0$, as is the case for the photonic lattice system discussed in the next section, then the two phases are separated by a gap-closing transition when $\eta = \pm\Omega/2$, despite the fact that the Chern number is the same on both sides of the transition.  This scenario evokes the mean-field phase diagram of the 2D Ising model, which exhibits a phase transition only in the absence of an external magnetic field [see Fig.~\ref{fig: phase diagrams} (b)].

We close this section by observing that the two phases to the left ($p_{\rm min}>0$) and right ($p_{\rm min}=0$) of the dashed line in Fig.~\ref{fig: phase diagrams}(b) can be distinguished experimentally by any measurement that is sensitive to the density of states $g(E)$.  In particular, one can show that the density of states diverges at the bottom of the conduction band (the basin depicted in Fig.~\ref{bands}) in the region of the phase diagram where $p_{\rm min}>0$.  More precisely, we find to leading order that
\begin{align}
g(E\to E_{\rm g}/2)\sim\frac{1}{\sqrt{E-\frac{E_{\rm g}}{2}}}.
\end{align}
If instead $p_{\rm min}=0$, then we find that $g(E\to E_{\rm g}/2)=\text{const.}$ to leading order.  One can use this difference in leading behavior near $E_{\rm g}/2$ to calculate the phase boundary indicated by the dashed line in Fig.~\ref{fig: phase diagrams}(b).  We find the parabolic phase boundary
\begin{align}
\abs{\frac{2\eta}{\Omega}} = 2\(\frac{\alpha}{\Omega}\)^2+1.
\end{align}

\section{Experimental Realizations}
\subsection{Graphene}
Before moving on to a proposal for implementing the above physics in photonic lattices, we first comment briefly on the prospects for realizing the same physics in graphene.  We focus on the case of the onsite potential, which is the dominant effect in graphene.  Realizing an onsite potential of the kind discussed in Sec.~III.A.~requires two key ingredients.  The first is the ability to excite the LO and LA phonon modes at high momentum ($\bm k=\bm K_+$, say).  These momenta can be accessed via surface physics techniques such as Helium scattering.\cite{kohnanomaly}  The second ingredient is the ability to excite the two sublattices of graphene with unequal amplitudes, which is necessary in order to open the gap predicted in Eq.~\eqref{onsitegap}.  This is a nontrivial feat, as bulk graphene has an intrinsic sublattice symmetry.  It is possible that this sublattice symmetry could be broken simply due to details of the edge termination of a given graphene flake.  A more interesting possibility is that the presence of the driving itself could break this symmetry dynamically.  This possibility is currently under investigation.

\subsection{Photonic Lattices}
The recent realization of graphene-like physics in photonic lattices provides a promising avenue for exploring the phase diagram of the $d$-wave model defined in Sec.~III.B.  The massless Dirac physics of graphene was demonstrated to exist in a quasi-two-dimensional system of evanescently coupled waveguides.\cite{peleg}  The waveguides are arranged in a honeycomb pattern such that the axis of propagation aligns with the $z$-axis.  The wave equation describing the paraxial propagation of light through the waveguide array can be written within coupled mode theory \cite{coupledmode} as
\begin{align}
i\partial_z\Psi_n=\sum_{m\in \text{NN}}c_{nm}(z)\Psi_m+\sum_{m\in \text{NNN}}c_{nm}(z)\Psi_m,
\end{align}
where $\Psi_n$ is the mode amplitude in waveguide $n$ and the matrix elements $c_{nm}$ depend on the overlap integral between the modes in waveguides $n$ and $m$.  The stationary modes of the waveguide system can thus be thought of as solutions of a Schr\"odinger equation in the tight-binding approximation, with time replaced by the propagation distance along the $z$-direction.  

Recently, a similar apparatus was used to realize a photonic system that is gapped in the bulk but exhibits topologically protected chiral edge modes.\cite{rechtsman}  In this set-up, helical waveguides were used to generate an effective $z$- (or time-) dependent gauge field that mimics the effect of circularly-polarized light on an electronic system.  However, in light of the analogy with tight-binding models, one can also think of a helical waveguide array as a crystal lattice whose sites rotate in time around their equilibrium positions.  As such, it is natural to propose the possibility of using such a waveguide array to realize the model discussed in this paper.  Indeed, the mode vectors \eqref{modes} map to helices parameterized as follows:
\begin{subequations}
\begin{align}
\bm r^A_{i}(s)&=\[R \cos(s-\phi_i),\;R \sin(s-\phi_i),\;\frac{Z}{2\pi}s\]\\
\bm r^B_{i}(s)&=\[-\pri R \cos(s-\phi_i),\;\pri R \sin(s-\phi_i),\;\frac{Z}{2\pi}s\],
\end{align}
\end{subequations}
where $s$ is a dimensionless parameter and $\phi_i=2\pi(i-1)/3$ with $i\in\{1,2,3\}$ labeling the inequivalent atoms within the six-atom basis of the new unit cell (see Fig.~\ref{dfiber}).  The dimensionful quantities on the right-hand side are $R$ and $\pri R$, which are the radii of the helices making up sublattices $A$ and $B$, and $Z$, the pitch of the helix.  In realistic systems, one can fabricate such an array of helices with $R,\pri R\sim 10\mu\text{m}$ and $Z\sim 1\text{cm}$.\cite{rechtsman}

One potential difficulty arising from the difference in helix radius between the two sublattices is that the resulting difference in arc length along the two helices can lead to decoherence due to phase accumulation in the NN hoppings.  The total phase difference accumulated over the length of the array is $\Delta\phi=2\pi\Delta L/\lambda$, where $\Delta L$ is the difference in path length between the two helices and $\lambda$ is the wavelength of light.  This phase accumulates linearly in $z$ at a rate $\omega = \Delta\phi/Z$.  The effects of this phase accumulation can be mitigated by minimizing $\Delta\phi$ subject to the constraint of avoiding the gap-closing when $R=\pri R$.  When $\omega\ll 2\pi/Z$, one can show that the phase accumulation will not affect the existence of a band gap or change the Chern number.

This experimental set-up has several practical advantages over the corresponding one in graphene.  Firstly, it is possible to tune $R$ and $\pri R$ separately, i.e.\ the degree of sublattice symmetry breaking can be tuned by hand in order to explore the full phase diagram of the $d$-wave model.  Furthermore, the pitch $Z$, which is analogous to the rotation period $2\pi/\Omega$ of the phonon mode, is a tunable parameter, in contrast to the case in graphene, where $\Omega\sim 150$ meV is fixed.  Because the size of the gap grows with increasing $\Omega$ [c.f.\ Eq.\ \eqref{gap}], this means that one can achieve larger photonic gaps by decreasing the pitch.

\section{Conclusion}
To summarize, in this work we have shown that it is possible to generate gaps in graphene-like systems that cannot be interpreted as mass gaps in the usual sense.  We illustrated this point by considering a dynamical coupling of the two Dirac points arising from the excitation of a superposition of phonon modes that modulate the two triangular sublattices independently.  This dynamical coupling leads to terms in the Hamiltonian that do not satisfy the anticommutation relations that define the usual Dirac masses.  We studied the transport properties of the system in the presence of these lattice modulations by making use of an exact mapping to a time-independent Hamiltonian that preserves all electric response functions, including conductivities.  We found that the system exhibits Chern insulating phases characterized by Hall conductivities $\sigma_{xy}=\pm e^2/h$, where the sign of the Chern number depends on both the momentum of the original phonon excitation ($\bm K_+$ or $\bm K_-$), as well as on which of sublattice $A$ or $B$ is excited with larger amplitude.  We further showed that the resulting model can be continuously deformed into the Haldane model unless the onsite potential modulation is suppressed, in which case the two models are separated by a gap-closing transition.  Finally, we proposed a scheme for realizing the physics studied in this paper in photonic lattices.

Several natural directions for future work present themselves.  For example, one can ask how to construct a revised classification of gapped Dirac Hamiltonians given that higher angular momentum structures are possible.  One can also consider adding more on-site degrees of freedom (e.g.\ spin and superconductivity) to the tight-binding Hamiltonian, so that higher-dimensional representations of the Dirac equation are realized and more matrix structures become possible.  Another worthwhile avenue is to consider in more detail the gap-closing that separates the $d$-wave and Haldane phases in the absence of an onsite potential modulation.  For example, can the band degeneracy at the critical point $(|2\eta/\Omega|,\alpha)=(1,0)$ be lifted by interactions?  Finally, the prospect of using photonic lattices to study this and other topological phases arising from dynamical lattice modulations is very intriguing, and future theory work could devise further novel applications of this approach.

\section*{Acknowledgments}
We are extremely grateful to Fernando de Juan, who pointed out to us the issue of the onsite potential modulation in graphene.  We also thank Emil Bergholtz, \'Alvaro G\'omez-Le\'on, and Takashi Oka for helpful discussions.  This work is supported in part by DOE Grant DEF-06ER46316 (T.I. and C.C.).

\bibliography{nnn_paper}
\end{document}